\documentclass[english,12pt,aps,prd,a4paper,preprintnumbers,floatfix,nofootinbib,showpacs,superscriptaddress, notitlepage,showkeys,amsmath,amssymb,nobibnotes]{revtex4-1}
\usepackage[colorlinks=true,citecolor=darkred,urlcolor=darkred, pdfborder={0 0 0}]{hyperref}
\usepackage{caption}
\usepackage{subcaption}
\usepackage[T1]{fontenc} 
\usepackage{subcaption}
\usepackage{amssymb}
\usepackage{hyperref}
\usepackage{bm}
\usepackage{verbatim}
\usepackage{setspace}
\usepackage{float}
\usepackage[usenames,dvipsnames]{xcolor}
\usepackage[colorinlistoftodos]{todonotes}
\usepackage[normalem]{ulem}

\usepackage[compat=1.1.0]{tikz-feynhand}
\usepackage{tikz-feynman}
\tikzfeynmanset{compat=1.1.0}
\usepackage{feynmp}
\usepackage{tikzsymbols}
\usepackage{array}
\usepackage{pifont} 
\usepackage{bbold}
\usepackage{pifont}
\usepackage{array}
\newcolumntype{L}[1]{>{\raggedright\let\newline\\\arraybackslash\hspace{0pt}}m{#1}}
\newcolumntype{C}[1]{>{\centering\let\newline\\\arraybackslash\hspace{0pt}}m{#1}}
\newcolumntype{R}[1]{>{\raggedleft\let\newline\\\arraybackslash\hspace{0pt}}m{#1}}

\definecolor{titlecolor}{rgb}{0.6,0,0}
\definecolor{mightnightblue}{RGB}{25,25,112}
\definecolor{brown}{rgb}{0.59, 0.29, 0.0}
\definecolor{darkred}{rgb}{0.6,0,0}
\definecolor{linkcolor}{rgb}{0,0,0.5}

\usepackage{natbib}

 \newcommand{\AddrIISERB}{Department of Physics,
 Indian Institute of Science Education and Research - Bhopal \\
 Bhopal Bypass Road, Bhauri, Bhopal, India}

\begin{document} 
\title{\boldmath \color{BrickRed} The Triplet Dirac Seesaw in the View of the Recent CDF-II W Mass Anomaly}
\author{Oleg Popov}
\email{opopo001@ucr.edu}
\affiliation{%
 Department of Physics, Korea Advanced Institute of Science and Technology, 291 Daehak-ro, Yuseong-gu, Daejeon 34141, Republic of Korea
}%
\author{Rahul Srivastava}
\email{rahul@iiserb.ac.in}
\affiliation{\AddrIISERB}
\date{\today}

\begin{abstract}
In the present letter, a Dirac neutrino mass model is presented in the view of the new result on W boson mass of $m_W^{CDF-II}=80.4335\pm 0.0094$ GeV, recently reported by the CDF-II experimental collaboration. The newly measured value of the W mass anomaly shows a 7-$\sigma$ deviation from that predicted by the standard model. The model explains the CDF-II W boson mass anomaly by extending the standard model with hypercharge zero vector-like fermion triplet. Symmetry is amended with a global $U(1)_{B-L}$ which is broken with an hypercharge zero electroweak triplet. It is demonstrated that the model can successfully explain CDF-II result, Dirac neutrino mass origin, while satisfying standard model precision constraints and collider constraints with scalar and fermion triplet masses in the ranger $1.4$ TeV$-5.2$ TeV and $100$ TeV $-10^{13}$ GeV, respectively.
\end{abstract}
\maketitle
%

%

\section{Introduction}
\label{sec:intro}

Standard Model (SM) of particle physics has been a highly successful theory with its predictions tested in many different experiments over several decades. The discovery of  125 GeV ``SM Higgs like'' scalar particle at Large Hadron Collider (LHC)  \cite{Aad:2012tfa,Chatrchyan:2012xdj} seems to complete the SM. However, there is a general expectation that SM cannot be the final theory of nature. 
One of the main experimental motivation for Beyond Standard Model (BSM) physics comes from the discovery of neutrino oscillations \cite{Super-Kamiokande:1998kpq}.  For neutrinos to oscillate from one flavour to another, neutrinos should have small but tiny masses. Since in SM neutrinos are massless, new physics is necessarily needed to generate small non-zero neutrino masses which can explain neutrino oscillations. In addition, the new result from Collider Detector at Fermilab (CDF) Collaboration taken at Tevatron particle accelerator also indicate that the W boson mass is 
7-$\sigma$ away from the SM predicted value~\cite{CDF:2022hxs} again implying the presence of new physics~\cite{Fan:2022dck,Lu:2022bgw,Athron:2022qpo,Yuan:2022cpw,Strumia:2022qkt,Yang:2022gvz,deBlas:2022hdk,Zhu:2022tpr,Du:2022pbp,Tang:2022pxh,Cacciapaglia:2022xih,Blennow:2022yfm,Sakurai:2022hwh,Fan:2022yly,Liu:2022jdq,Lee:2022nqz,Cheng:2022jyi,Bagnaschi:2022whn,Paul:2022dds,Bahl:2022xzi,Asadi:2022xiy,DiLuzio:2022xns,Athron:2022isz,Gu:2022htv,Heckman:2022the,Babu:2022pdn,Zhu:2022scj,Balkin:2022glu,Biekotter:2022abc,Endo:2022kiw,Crivellin:2022fdf,Cheung:2022zsb,Du:2022brr,Heo:2022dey,Krasnikov:2022xsi,Ahn:2022xeq,Han:2022juu,Zheng:2022irz,Perez:2022uil,Ghoshal:2022vzo,Kawamura:2022uft,Nagao:2022oin,Kanemura:2022ahw}. In this work we aim to provide a coherent new physics description which can simultaneously explains the CDF W-boson anomaly via physics related to neutrino mass generation. 

Coming back to possible new physics required to explain CDF anomaly along with generating neutrino mass, one should ask the question about nature of neutrinos i.e. whether they are Dirac or Majorana particles. Experimentally both possibilities are still open as experiments such as neutrinoless double beta decay experiments have not yet seen any signature for Majorana neutrinos \cite{KamLAND-Zen:2022tow}. Theoretically, in past most of the research in neutrino physics was focused on Majorana neutrinos \cite{Minkowski:1977sc,Yanagida:1979ab,GellMann:1979grs, Mohapatra:1979ia,Schechter:1980gr,Foot:1988aq,Zee:1980ai,Babu:1988ki,Ma:2006km,Fraser:2014yha}, however in recent times mass models for Dirac neutrinos have garnered quite a bit of attention  \cite{Ma:2014qra,Ma:2015raa,Ma:2015mjd, CentellesChulia:2016rms, CentellesChulia:2016fxr, Ma:2016mwh, CentellesChulia:2017koy, CentellesChulia:2018gwr, CentellesChulia:2018bkz, Bonilla:2018ynb, CentellesChulia:2019xky, Srivastava:2019xhh, Peinado:2019mrn,Kang:2018lyy,Dasgupta:2019rmf,Dasgupta:2021ggp}. In this work we will look at the possibility of Dirac neutrino mass models which can also explain the CDF anomaly.  

Before moving on to explain the mass models  of neutrino physics one should also consider what kind of neutrino mass models can potentially explain the CDF anomaly. Since, the anomaly implies that the W boson mass is larger than the SM prediction\footnote{The numerical SM prediction for W-boson mass is obtained using the experimental values of the precision observable and the SM relationship between them and the W-boson.}. This anomalous behaviour can only be explained by some new physics effecting the W boson mass at tree or loop level. Given the fact that the anomaly is in W boson mass, one can immediately draw some conclusions on the type of new physics required to explain the anomaly. The new physics should be such that it selectively changes the W boson mass but should have minimal or no correction to Z boson mass as neither CDF nor any other experiment has seen any anomaly in Z boson mass. This fact implies that purely neutral particles such as heavy right handed neutrinos will play little to no role in explaining the anomaly, thus restricting the type of neutrino mass models, as we discuss in next section.

The paper is organized as follows: In Section \ref{sec:neutrino} we discuss the different Dirac  mass models using the effective operator approach.
We identify the  Dirac triplet seesaw as the most promising simple candidate. In Section \ref{sec:model} we provide details of the triplet seesaw model and work out the mass spectrum of the particles in the model. In Section \ref{sec:neutrino_mass} we show the parameter space of the model which leads to correct neutrino masses consistent with the various neutrino experiments. In Section \ref{sec:gauge_mass} we work out the corrections to the gauge boson masses in our model. In Section \ref{sec:analysis_constraints} we present our results taking into account all the constraints on our model. Finally, we conclude in Section \ref{sec:conclusion}.

\section{Dirac neutrino mass models}
\label{sec:neutrino}

The nature of neutrinos i.e. whether they are Majorana or Dirac particles is still an open question in neutrino physics. The oscillation experiments are insensitive to the Dirac/Majorana nature of neutrinos while the experiments sensitive to neutrino nature such as neutrinoless double beta decay experiments have not seen any signals so far \cite{KamLAND-Zen:2022tow}. Traditionally, neutrinos were assumed to be Majorana particles and many different mass mechanisms for them were developed. However, recently the Dirac neutrino mass models have received considerable attention with several novel Dirac neutrino mass mechanism have been developed. Furthermore, several operator based analysis for Dirac neutrion mass models have also been considered in recent past \cite{Ma:2016mwh, CentellesChulia:2018gwr, CentellesChulia:2018bkz, Bonilla:2018ynb, CentellesChulia:2019xky}. To have a successful mass mechanism for Dirac neutrinos several key conditions need to be satisfied:
\begin{itemize}
 \item If neutrinos are Dirac then right handed neutrinos $\bar{\nu}$ need to be added to SM particle content\footnote{We follow the notation where all fermions are left handed. Hence, instead of calling the right handed neutrinos $\nu_R$ we define its charge conjugate which will be  a left handed field.}
 \item An unbroken symmetry $S$ under which neutrinos transform non-trivially,  is needed to ensure that neutrino remain Dirac particles to all quantum loops.
 \item Tree level coupling between right and left handed neutrinos needs to be forbidden.
\end{itemize}
Assuming these conditions one can write down the effective operator which will generate the Dirac neutrino masses. The lowest dimensional operator invariant under SM  $SU(3)_C \otimes SU(2)_L \otimes U(1)_Y$ gauge symmetries and the $S$ symmetry is given by
\begin{eqnarray}
 \frac{\kappa_{ij}}{\Lambda}\, L_i X_1 X_2 \bar{\nu}_j
 \label{dim-op}
\end{eqnarray}
where $\kappa_{ij}$ are the dimensionaless coupling constants and $\Lambda$ is the cutoff scale. Also, $L_i$; $i = 1,2,3$ are the SM lepton doublets, $\bar{\nu}_j$; $j = 1, \cdots N$ are the SM gauge singlet right handed neutrinos and $X_1, X_2$ are VEV carrying scalars whose SM charges should be such that \eqref{dim-op} remains invariant under the SM gauge group. There is no constraints on the maximum number of right handed neutrinos, however the neutrino oscillation data implies that at least two neutrinos should be massive, which in turn means that there should be atleast two right handed neutrinos. Furthermore, either $L$ or $\bar{\nu}$ or both should be charged under the unbroken symmetry $S$ such that Majorana mass terms to all loops are forbidden. To ensure that $S$ remains unbroken, neither of the two scalars should carry any non-trivial charges under the $S$ symmetry. Finally, either the $S$ symmetry itself or some other mechanism should ensure that the direct coupling between $L$ and $\bar{\nu}$ through SM Higgs doublet $H$ is forbidden.

If one chooses one of the scalars, let's say $X_1$ in \eqref{dim-op}, to be the Higgs doublet $H$ i.e. $X_1 \equiv H$, then the $SU(2)_L$ gauge invariance immediately fixes the other scalar $X_2$ to be either singlet or triplet under the $SU(2)_L$ symmetry. The hypercharge of the scalar $X_2$ can also be immediately fixed to be $Y = 0$ by the demand that the operator in \eqref{dim-op} remains invariant under the $U(1)_Y$ gauge symmetry. Needless to say that none of the scalars $X_1, X_2$ should carry any non-trivial charge under the $SU(3)_C$ color symmetry. With the above conditions the two possible dim-5 operators are 
\begin{eqnarray}
 \frac{\kappa_{ij}}{\Lambda}\, L_i H \sigma \bar{\nu}_j  \nonumber \\
 \frac{\kappa_{ij}}{\Lambda}\, L_i H \Delta \bar{\nu}_j 
 \label{dim5-op}
\end{eqnarray}
where $\sigma$ is the $SU(2)_L$ singlet scalar while $\Delta$ transforms as a triplet under the $SU(2)_L$ symmetry. The possible Ultra Violet (UV) completions of these operators have been discussed in \cite{Ma:2016mwh, CentellesChulia:2018gwr, CentellesChulia:2018bkz}. 

Before going any further with the details of the possible UV completions, let's first discuss the nature and possible examples of the $S$ symmetry which is required to ensure Diracness of the neutrinos.
There are many options for the $S$ symmetry which can be a simple abelian discreet symmetry \cite{Ma:2014qra,Ma:2015raa,Ma:2015mjd,CentellesChulia:2016rms} or can be a non-abelian discreet symmetry \cite{CentellesChulia:2016fxr}, even modular symmetry~\cite{Dasgupta:2021ggp}. Continuous symmetries also can be employed \cite{Peinado:2019mrn}. The simplest option for $S$ seems to be to use the unbroken subgroups of the Lepton number $U(1)_L$ symmetry \cite{Ma:2014qra,Ma:2015raa,Ma:2015mjd, Hirsch:2017col,Bonilla:2018ynb,Srivastava:2019xhh}. In this paper we will take this route.

Since, the Lepton number $U(1)_L$ symmetry is anomalous with SM particle content, it is better to consider a linear combination of it with the Baryon number $U(1)_B$ symmetry. The resulting $U(1)_{B-L}$ symmetry can be made anomaly free by addition of three right handed neutrinos. There are two possible solutions namely
\begin{itemize}
 \item \textbf{Vector Solution:} The three right handed neutrinos carry charges $\bar{\nu}_i \sim (-1,-1,-1)$; $i = 1,2,3$ under $B-L$ symmetry.
 \item \textbf{Chiral Solution:} The three right handed neutrinos carry charges $\bar{\nu}_i \sim (-4,-4, +5)$; $i = 1,2,3$ under $B-L$ symmetry.
\end{itemize}

In past the vector solution was often used to generate Majorana neutrino masses through the explicit or spontaneous (through a scalar carrying two units of $B-L$ charge) breaking $U(1)_{B-L} \to \mathbb{Z}_2$. However, the chiral solution naturally leads to Dirac neutrino mass models \cite{Ma:2014qra,Ma:2015raa,Ma:2015mjd} with $U(1)_{B-L} \to \mathbb{Z}_3$ explicit or spontaneous breaking. The resulting unbroken $\mathbb{Z}_3$ symmetry can then play the role of the $S$ symmetry and can ensure Dirac nature of neutrinos to all loop orders. Further advantage of using the chiral solution is that the tree level direct coupling between $L$ and $\bar{\nu}$ is automatically forbidden. Finally, being a anomaly free symmetry, $U(1)_{B-L}$ can be trivially gauged leading to a richer phenomenology. There are many possible UV completions of the operators in \eqref{dim5-op} both at tree and loop level. In this work, keeping CDF anomaly in mind, we will concentrate on a particular solution, namely the Dirac Triplet seesaw model, which we now discuss in details in next section.

\section{The Dirac Triplet Seesaw Model}
\label{sec:model}

We now focus on a particular seesaw completion of the operator $ \frac{\kappa_{ij}}{\Lambda}\, L_i H \Delta \bar{\nu}_j$ which we call the Dirac triplet seesaw model. The particle content and  their charges under $SU(3)_C \otimes SU(2)_L \otimes U(1)_Y$ gauge charges as well as their charges under the global $U(1)_{B-L}$ symmetry are given in Table \ref{tab:particle_content}.

\begin{table}[!h]
    \centering
    \begin{tabular}{| c | c | c | c | c |}
        \hline  
& \hspace{0.5cm} $SU(3)_C$  \hspace{0.5cm} &   \hspace{0.5cm} $SU(2)_L$   \hspace{0.5cm} &   \hspace{0.5cm} $U(1)_Y$   \hspace{0.5cm} &   \hspace{0.5cm} $U(1)_{B-L}$   \hspace{0.5cm} \\ \hline \hline
        $Q$ & $\pmb{3}$ & $\pmb{2}$ & $\frac{1}{6}$ & $\frac{1}{3}$ \\
        $\Bar{u}$ & $\pmb{\Bar{3}}$ & $\pmb{1}$ & $-\frac{2}{3}$ & $-\frac{1}{3}$ \\
        $\Bar{d}$ & $\pmb{3}$ & $\pmb{1}$ & $\frac{1}{3}$ & $-\frac{1}{3}$ \\
        $L$ & $\pmb{1}$ & $\pmb{2}$ & $-\frac{1}{2}$ & $-1$ \\
        $\Bar{e}$ & $\pmb{1}$ & $\pmb{1}$ & $1$ & $1$ \\
        $\Bar{\nu}$ & $\pmb{1}$ & $\pmb{1}$ & $0$ & $4,4,-5$ \\
        $\Sigma, \Bar{\Sigma}^{\dagger}$ & $\pmb{1}$ & $\pmb{3}$ & $0$ & $1$ \\ \hline
        $H$ & $\pmb{1}$ & $\pmb{2}$ & $\frac{1}{2}$ & $0$ \\
        $\Delta$ & $\pmb{1}$ & $\pmb{3}$ & $0$ & $-3$ \\ \hline 
    \end{tabular}
    \caption{Model field content and their charges under symmetries of our model. For brevity we have suppressed the flavour indices, see text for details. }
    \label{tab:particle_content}
\end{table}

As can be seen from Tab.\ref{tab:particle_content}, apart from SM particles we have added three right handed neutrinos which are singlet under SM gauge group but carry $(-4,-4,+5)$ charges under the global $U(1)_{B-L}$ symmetry. In addition we have added two pair of fermions $\Sigma, \bar{\Sigma}$ and a new scalar $\Delta$. Both $\Sigma$ and $\Delta$ carry no color charge, transform as triplets under $SU(2)_L$ and have zero charge under $U(1)_Y$ symmetry. Furthermore, their $U(1)_{B-L}$ charges are uniquely fixed by the Dirac seesaw mechanism as we discuss in section \ref{sec:neutrino_mass}.

With the particle content in Tab.\ref{tab:particle_content} the new Yukawas and total scalar potential Lagrangian are given by

\begin{align}
    \label{eq:yuk_lag}
    \mathcal{L} &= - Y_L L \Sigma H - Y_R \Bar{\nu} Tr\left[\Bar{\Sigma} \Delta \right] - M_\Sigma \Sigma \Bar{\Sigma} - V + \text{h.c.}, \\
    \label{eq:potential}
    V &= m_h^2 H^\dagger H + \frac{\lambda_h}{2} \left(H^\dagger H\right)^2 + \lambda_{h\Delta} \left(H^\dagger H\right) Tr\left(\Delta^\dagger \Delta\right) + \lambda_{h\Delta}^\prime \left(H^\dagger \Delta \Delta^\dagger H\right) \\
    &+ m_\Delta^2 Tr\left(\Delta^\dagger \Delta\right) + \frac{\lambda_\Delta}{2} Tr\left(\Delta^\dagger \Delta\right)^2 + \lambda_\Delta^\prime Tr\left(\Delta^\dagger \Delta \Delta^\dagger \Delta \right) + \mu_H H^\dagger \Delta H. \nonumber
\end{align}

As can be seen from \eqref{eq:yuk_lag} the $SU(2)_L$ triplet fermions couple with both $L$ and $\bar{\nu}$ through Yukawa couplings involving the doublet scalar $H$ and the triplet scalar $\Delta$, respectively. In addition to the Yukawa couplings, the $\Sigma, \bar{\Sigma}$ also have an invariant mass term $M_\Sigma$ which being independent of the electroweak scale, can be much larger than the 
vacuum expectation value (VEV) of the scalars i.e. $M_\Sigma >> v_h,v_\Delta$. Furthermore, the presence of the cubic term $\mu_H H^\dagger \Delta H$ in \eqref{eq:potential} implies that the global $U(1)_{B-L}$ symmetry is explicitly broken and the nambu$-$goldstone boson gets a mass proportions to $\mu_H$ parameter. Additionally, the mass splitting of the neutral and charged components of the triplet scalar is also proportional to $v_\Delta \mu_H$. This leads to strong constraints on scalar masses as discuss in sec.~\ref{sec:STU_parameters}.

\subsection{Neutrino Masses}
\label{sec:neutrino_mass}

\begin{figure}[!h]
        \centering
		\includegraphics[width=0.6\textwidth, trim=8cm 24cm 7cm 2cm, clip]{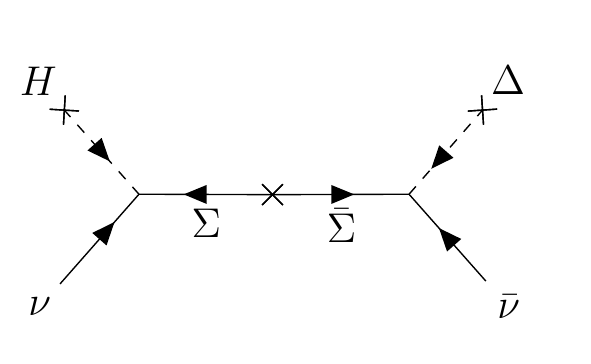}
        \caption{Tree level Dirac seesaw neutrino mass Feynman diagram. The charge of $\Delta$ under $U(1)_{B-L}$ is $-3$.}
        \label{fig:neutrino_dia_1}
\end{figure}

The Feynman diagram for the seesaw mass for the neutrinos is given in Fig.\ref{fig:neutrino_dia_1}

Since, the $SU(2)_L$ triplet scalar $\Delta$ transform as $-3$ under the $U(1)_{B-L}$ symmetry, its VEV spontaneously breaks $U(1)_{B-L} \to \mathbb{Z}_3$ residual subgroup. This $\mathbb{Z}_3$ remains unbroken forbidding the Majorana mass terms for neutrinos ensuring the Dirac nature of neutrinos.

The Dirac mass matrix for the neutral fermions is given by

\begin{align}
    \label{eq:neutral_fermion_m}
  M_{\nu\Sigma} \, = \,  &\left(
    \begin{matrix}
    \nu & \Bar{\Sigma}
    \end{matrix}
    \right) \left(
    \begin{matrix}
    0 & Y_L \frac{v_h}{\sqrt{2}} \\ Y_R \frac{v_\Delta}{\sqrt{2}} & M_{\Sigma}
    \end{matrix}
    \right) \left(
    \begin{matrix}
    \Bar{\nu} \\ \Sigma
    \end{matrix}
    \right),
\end{align}

in the $(\nu, \Bar{\Sigma})$ (from left) and $(\Bar{\nu}, \Sigma)$ (from right) basis, while all Majorana mass terms are forbidden by the residual $\mathbb{Z}_3$  discrete symmetry~\cite{Ma:2016mwh}. Neutrino and the heavy Dirac masses are given by the Dirac seesaw mechanism

\begin{subequations}
\label{eq:neutral_m}
\begin{align}
    \label{eq:nu_mass}
    m_\nu &\simeq - Y_L Y_R \frac{v_h v_\Delta}{M_\Sigma},\\
    \label{eq:heavy_mass}
    m_\Sigma &\simeq M_\Sigma + Y_L Y_R \frac{v_h v_\Delta}{M_\Sigma},
\end{align}
\end{subequations}

where as the left and right mixing angles are given by

\begin{subequations}
\label{eq:mix_angles}
\begin{align}
    \label{eq:left_mix_angle}
    \tan\left(2 \theta_L \right) = \frac{2 \sqrt{2} Y_L v_h M_\Sigma}{ Y_L^2 v_h^2 - Y_R^2 v_\Delta^2 - 2 M_\Sigma^2}, \\
    \label{eq:right_mix_angle}
    \tan\left(2 \theta_R \right) = \frac{2 \sqrt{2} Y_R v_\Delta M_\Sigma}{ Y_R^2 v_\Delta^2 - Y_L^2 v_h^2 - 2 M_\Sigma^2}.
\end{align}
\end{subequations}

Before ending let us note an interesting feature of our model. As we will discuss shortly, various constraints imply that the VEV of the triplet scalar $v_\Delta$ should be far smaller than the VEV of doublet scalar $v_h$ i.e. $v_\Delta << v_h$. This means that the seesaw mass of light neutrinos in \eqref{eq:nu_mass} is actually doubly suppressed both by smallness of $v_\Delta$ as well as heaviness of $M_\Sigma$. This is actually the inverse seesaw limit, an interesting feature of a wide variety of Dirac seesaw models as discussed in \cite{CentellesChulia:2020dfh}.
%

\subsection{Gauge Boson Masses}
\label{sec:gauge_mass}

The addition of the scalar triplet $\Delta$ has immediate consequences for the mass of the gauge bosons. The covariant derivative Lagrangian for scalars is given by 

\begin{align}
    \label{eq:gauge_kin}
    \mathcal{L} &= \left(D_{\mu} H\right)^\dagger \left(D^\mu H\right) + Tr \left[ \left(D_\mu \Delta_{2\times 2} \right)^\dagger \left(D^\mu \Delta_{2\times 2} \right) \right],
\end{align}
where 
\begin{align}
    \label{eq:cov_der}
    D^\mu &= \partial^\mu + \imath g W^{\mu}_a \frac{\sigma^a}{2} + \imath g^\prime \frac{Y}{2} B^\mu,
\end{align}
and 
\begin{subequations}
\label{eq:vevs}
\begin{align}
    \label{eq:vev_h}
    H &= \frac{1}{\sqrt{2}} \left(\begin{matrix} H^\pm \\ H^0 \end{matrix}\right), &\left\langle H \right\rangle = \frac{1}{\sqrt{2}} \left(\begin{matrix} 0 \\ v_h \end{matrix}\right), \\
    \Delta &= \frac{1}{\sqrt{2}} \left( \begin{matrix} \Delta^+ \\ \Delta^0 \\ \Delta^- \end{matrix} \right), &\left\langle \Delta \right\rangle = \frac{1}{\sqrt{2}} \left( \begin{matrix} 0 \\ v_\Delta \\ 0 \end{matrix} \right), \\
    \Delta_{2\times 2} &= \left( \begin{matrix} \Delta^0 / \sqrt{2} & \Delta^+ \\ \Delta^- & -\Delta^0 / \sqrt{2} \end{matrix} \right), &\left\langle \Delta_{2\times 2} \right\rangle = \left\langle \Delta_a \right\rangle \left(\sigma^a\right)_{ij} = \frac{1}{\sqrt{2}} \left( \begin{matrix} v_\Delta & 0 \\ 0 & -v_\Delta \end{matrix} \right).
\end{align}
\end{subequations}

We take the EW doublet VEV, $v_h$ to be the SM VEV $v_{SM}$ ($\equiv 246.221$ GeV), and new the scalar triplet VEV, $v_\Delta$, as the BSM contribution, the sum is $v^2 = v_h^2 + 8 v_\Delta^2$. In this case, the SM gauge boson masses and Weinberg angle are given by

\begin{subequations}
\label{eq:gauge_masses}
\begin{align}
    M_\gamma &= 0, \\
    M_Z^2 &= \left( g^2+g^{\prime 2} \right) \frac{v_h^2}{4}, \\
    M_{W^\pm}^2 &= \frac{g^2}{4} \left( v_h^2 + 8 v_\Delta^2 \right), \\
    \cos \left( \theta_W \right) &= \frac{g}{\sqrt{g^2 + g^{\prime 2}}}, \\
    \sin \left( \theta_W \right) &= \frac{g^\prime}{\sqrt{g^2 + g^{\prime 2}}}.
\end{align}
\end{subequations}

As can be seen from \eqref{eq:gauge_masses} the abelian sector is left unaffected, therefore Weinberg angle and $Z$ boson mass remain as in SM. While the extra triplet contributes to the W mass and effects the $\rho$ parameter as will be shown next.

Custodial symmetry violation due to triplet's VEV is given by

\begin{align}
\label{eq:rho_oblique_param}
    \rho &\equiv \frac{m_W^2}{m_Z^2 \cos^2(\theta_W)} = 1 + 8\frac{v_\Delta^2}{v_h^2}.
\end{align}

The most recent PDG value of $\rho_{exp} = 1.00038 \pm 0.00020$~\cite{Zyla:2020zbs} has to be scaled by a factor of $\left(m_W^{CDF}/m_W^{PDG}\right)^2$ to give updated value of $\rho \simeq 1.00219\pm 0.00044$ after including the CDF results. Using \eqref{eq:rho_oblique_param} gives $v_\Delta \simeq 4.06997 \pm 0.26529$~GeV, which in turn is consistent with the triplet VEV of Fig.~\ref{fig:w_mass_anomaly}, to be obtained directly from the $m_W^{CDF}$ value fit in Sec.~\ref{sec:w_mass_anomaly}.
%

\section{Analysis and Constraints}
\label{sec:analysis_constraints}

Below we discuss the relevant constraints on the model, including neutrino mass, EW precision, and current and future collider constraints.
%

\subsection{CDF II W mass anomaly}
\label{sec:w_mass_anomaly}

\begin{figure}[!h]
    \centering
    \includegraphics[width=0.7\textwidth]{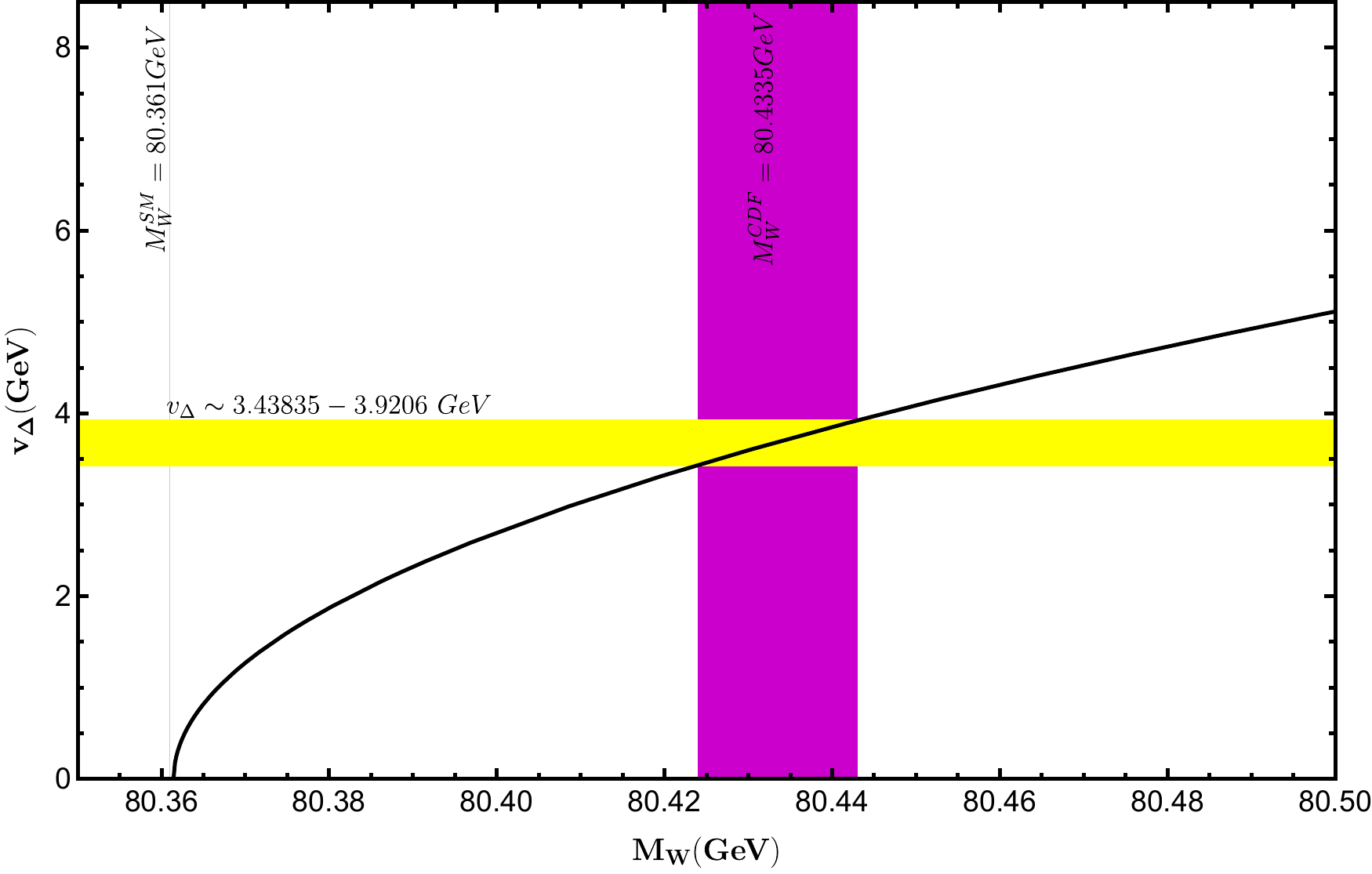}
    \caption{EW triplet VEV constraints due to recently reported CDF II W mass anomaly~\cite{CDF:2022hxs}.}
    \label{fig:w_mass_anomaly}
\end{figure}

Recent reported result by CDF II collaboration differs from SM value by 7-$\sigma$. In order to explain this deviation with a scalar triplet model $\left\langle \Delta \right\rangle \ll \left\langle H \right\rangle$. This fits naturally into the (Dirac) triplet seesaw mechanism. The range of valid EW triplet VEV is between $3.4384~GeV<\left\langle \Delta \right\rangle<3.9206~GeV$ and is shown in fig.~\ref{fig:w_mass_anomaly}.

%

\subsection{Neutrino mass constraints}
\label{sec:neutrino_mass_constraints}

\begin{figure}[!h]
    \centering
    \includegraphics[width=0.7\textwidth]{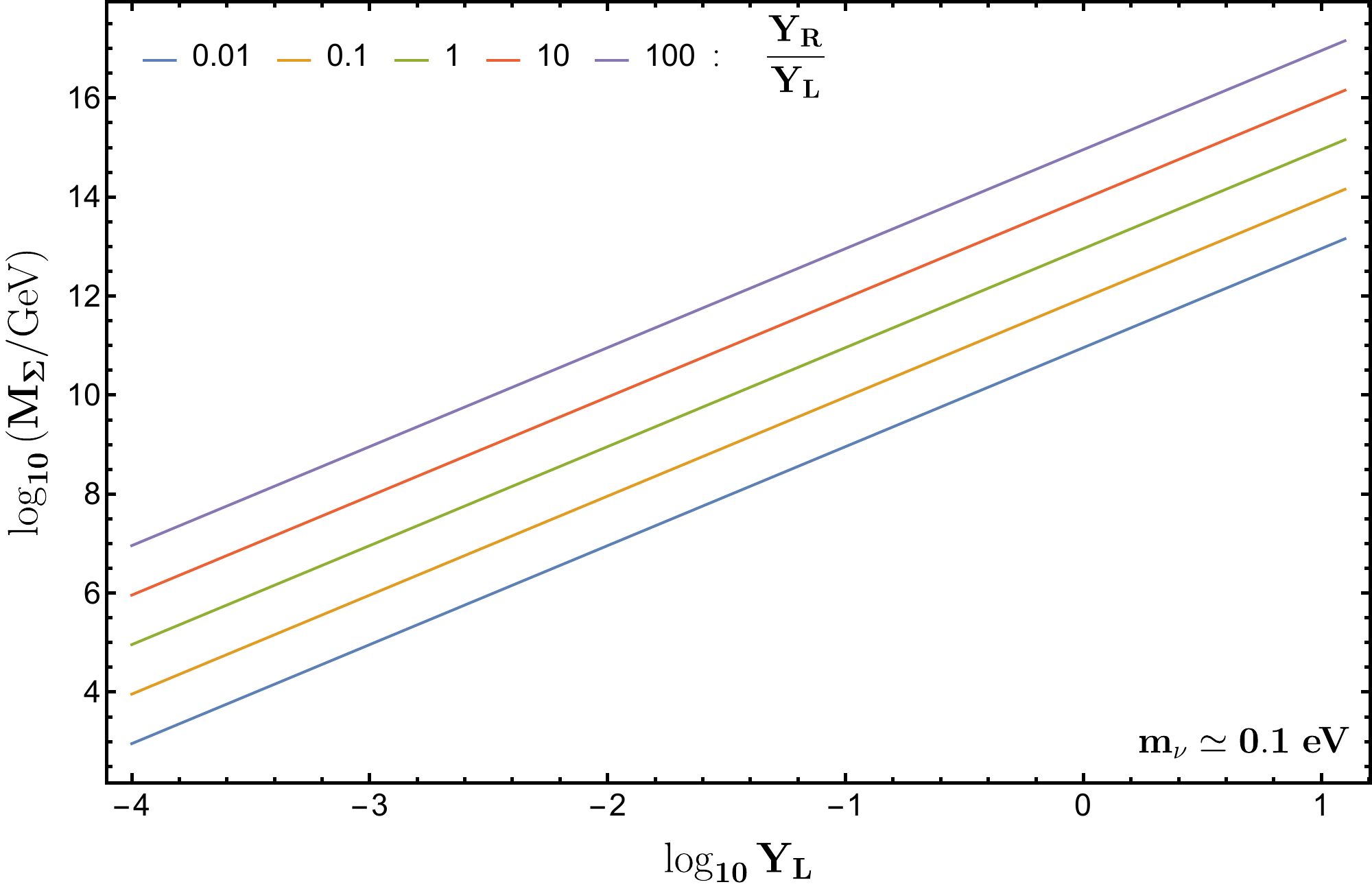}
    \caption{Correlation between $M_\Sigma$ and $Y_{L,R}$, for various ratios of $Y_L/Y_R$, that satisfy the observed neutrino mass.}
    \label{fig:m_nu_const}
\end{figure}

Dirac seesaw neutrino mass is given by \eqref{eq:nu_mass}. The natural way to satisfy the small neutrino masses is the combination of small scalar triplet VEV and large ($\gg \left\langle H\right\rangle$) fermion triplet mass. The correlation between relevant parameters is shown in Fig.~\ref{fig:m_nu_const}. As can be seen from the figure for the Yukawas of the order $\mathcal{O}\left(0.1\right)$ the fermion triplet must be of the order $\mathcal{O}\left(10^{10-11}\right)$~GeV to satisfy the neutrino mass constraint. On the other hand, if Yukawas are of the order $\mathcal{O}\left(10^{-4}\right)$ then the heavy fermion triplet can be as light as $\mathcal{O}\left(10-100\right)$~TeV, which brings it into a detectable range for future hadronic/leptonic colliders ($100$~TeV collider or FFC).

%

\subsection{Bounds from electroweak precision experiments}
\label{sec:STU_parameters}

The most important relevant electroweak precision constraints(EWPC) are the oblique, \emph{aka} Peskin$-$Takeuchi~\cite{Peskin:1991sw} S, T, U parameters~\cite{Funk:2011ad,Khan:2016sxm,Mandal:2022zmy}, which represent the contribution of BSM physics to the SM gauge sector. Since the triplet scalar carries no hypercharge, its contribution to $S$ parameter is absent. On the other hand, the triplet fermion contributes to S, T, U parameters only at the loop order. Furthermore, since the BSM EW triplet fermion is expected to be much heavier than the EW scale, its contribution to S, T, U oblique parameters is expected to be of subleading order. As a result of the above, new physics contribution to S, T, U parameters from scalar triplet with zero hypercharge is given by~\cite{Forshaw:2003kh,Forshaw:2001xq}

\begin{subequations}
\label{eq:stu_parameters}
\begin{align}
    \label{eq:s_param}
    S &\simeq 0, \\
    \label{eq:t_param}
    T &= \frac{1}{8\pi} \frac{1}{\sin^2 \theta_W \cos^2 \theta_W} \left[ \frac{M_{\Delta^0}^2 + M_{\Delta^\pm}^2}{M_Z^2} - \frac{2 M_{\Delta^0}^2 M_{\Delta^\pm}^2}{M_Z^2\left(M_{\Delta^0}^2 - M_{\Delta^\pm}^2\right)} \log \left(\frac{M_{\Delta^0}^2}{M_{\Delta^\pm}^2}\right) \right] , \nonumber \\
    &\simeq \frac{1}{8\pi} \frac{1}{\sin^2 \theta_W \cos^2 \theta_W} \frac{\left(\Delta M\right)^2}{M_Z^2}, \\
    \label{eq:u_param}
    U &= -\frac{1}{3\pi} \left( M_{\Delta^0}^4 \log\left(\frac{M_{\Delta^0}^2}{M_{\Delta^\pm}^2}\right) \frac{\left(3 M_{\Delta^\pm}^2 - M_{\Delta^0}^2 \right)}{\left(M_{\Delta^0}^2 - M_{\Delta^\pm}^2\right)^3} + \frac{5\left(M_{\Delta^0}^4 + M_{\Delta^\pm}^4\right) - 22 M_{\Delta^0}^2 M_{\Delta^\pm}^2}{6 \left(M_{\Delta^0}^2 - M_{\Delta^\pm}^2\right)^2}\right) \nonumber \\
    &\simeq \frac{\Delta M}{3\pi M_{\Delta^{\pm}}},
\end{align}
\end{subequations}

where $\Delta M = M_{\Delta^\pm} - M_{\Delta^0}$ and $\Delta M \ll M_{\Delta^\pm, \Delta^0}$ in the limit $M_{\Delta^\pm, \Delta^0} \gg M_h$. The current global fit of electroweak precision data (EWPD) after including the CDF result, gives~\cite{Lu:2022bgw}

\begin{subequations}
\label{eq:stu_exp_limits}
\begin{align}
    \label{eq:s_exp_limits}
    S = 0.06 \pm 0.10, \\
    \label{eq:t_exp_limits}
    T = 0.11 \pm 0.12, \\
    \label{eq:u_exp_limits}
    U = 0.13 \pm 0.09.
\end{align}
\end{subequations}

\begin{figure}[!h]
    \centering
    \includegraphics[width=0.6\textwidth]{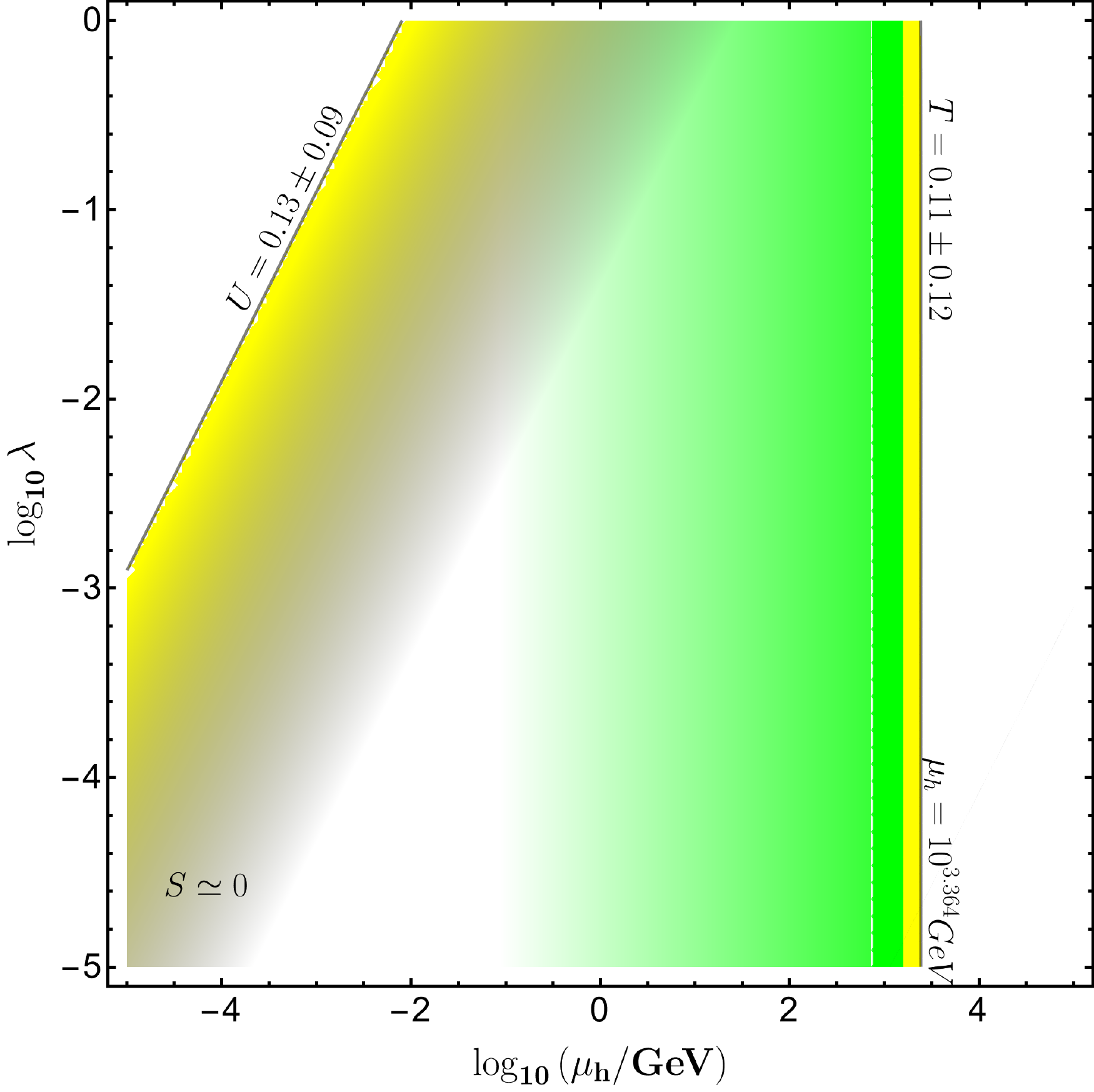}
    \caption{$T$ and $U$ Peskin$-$Takeuchi constraints on the model parameters. Green (yellow) shading corresponds to $1(2)$ standard deviation bounds from updated global SM fits~(eq.~\ref{eq:stu_exp_limits}). $S\simeq 0$ since there is no BSM physics contribution to the $Z-$boson. $\lambda_{H\Delta},\lambda_{H\Delta}^\prime \ll \lambda_\Delta = \lambda_\Delta^\prime = \lambda$ is taken here.}
    \label{fig:stu_plot}
\end{figure}

The oblique parameter constraints for the present model are shown in fig.~\ref{fig:stu_plot}. As can be seen from the plot, the most important constraint comes from T parameter and gives an upper limit on $\mu_H\simeq 2.3$~TeV parameter of the scalar potential. Further, the region of very small $\mu_H$ ($\leq 10$~MeV) is excluded for $10^{-3}<\lambda_\Delta = \lambda_\Delta^\prime = \lambda$ in the two sigma range. For the demonstration purposes we took $\lambda_{H\Delta},\lambda_{H\Delta}^\prime \ll \lambda_\Delta = \lambda_\Delta^\prime = \lambda$ in fig.~\ref{fig:stu_plot}.

%

\subsection{Collider Constraints}
\label{sec:collider_constraints}

\begin{figure}[!h]
    \centering
    \includegraphics[width=0.9\textwidth]{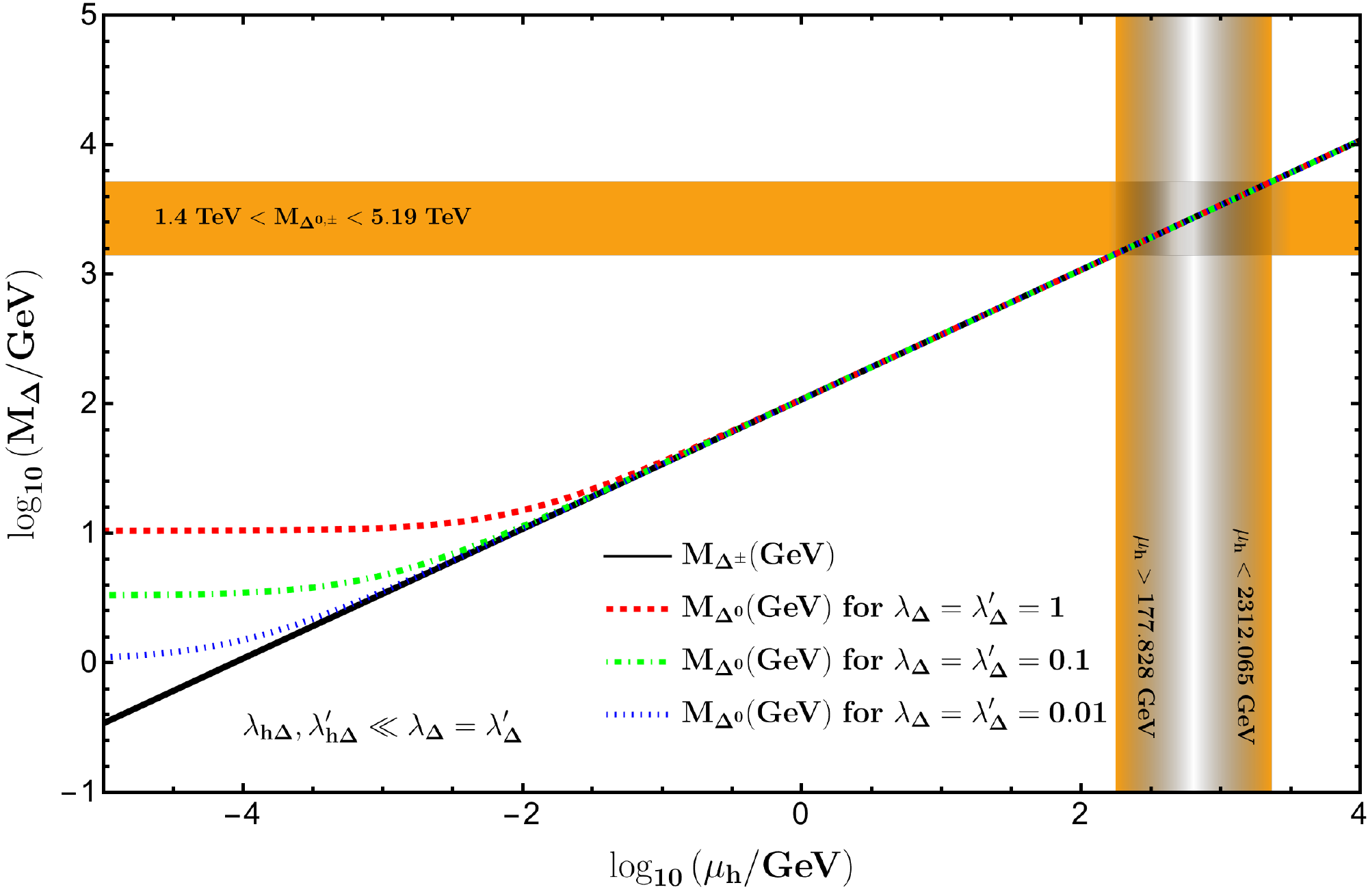}
    \caption{Mass constraints on the charged and neutral EW scalar triplet components. Upper limit is due to $T$ oblique parameter constraint, while the lower limit is due to LHC and FCC constraints.}
    \label{fig:delta_mass_limits}
\end{figure}

Since the model contains no BSM singlet particle, all new physics is subject to stringent collider constraints from LHC. Furthermore, the proposed future colliders such as FCC and future lepton collider can further probe the  EW scalar and fermionic triplets of our model. The most up$-$to$-$date limit on the fermion EW triplet comes from the CMS collaboration and is given by $M_\Sigma > 574~GeV$~\cite{CMS:2013czn}. Whereas, the current LHC limit on the charged component of the scalar EW triplet is given by $m_{\Delta^{\pm,0}} > 1065~GeV$~\cite{ATLAS:2018rns}. The proposed FCC can potentially improve the limits on scalar EW triplet upto  $m_{\Delta^{\pm,0}}>1.4~TeV$~\cite{Mandal:2022zmy}. The combination of the strongest collider bound and the oblique parameter constraints is shown in fig.~\ref{fig:delta_mass_limits}. As can be seen from this figure, the only valid parameter window for this model left is given by $177.8~GeV<\mu_H<2312~GeV$. Which translates into $1.4~TeV<m_{\Delta}<5.19~TeV$ bounds on scalar EW triplet for $\Delta m_{\Delta}\ll m_{\Delta^\pm}\approx m_{\Delta^0} \approx m_\Delta$ approximation, as can be seen from the Fig.~\ref{fig:delta_mass_limits}.
%

\section{Conclusion}
\label{sec:conclusion}

In the present letter, a model based on the scalar and fermion EW triplet was presented. The model is shown to successfully generate naturally small Dirac neutrino masses via a Dirac seesaw mechanism with a spontaneous symmetry breaking of the global $U(1)_{B-L}$ symmetry. Furthermore, the model successfully reproduces the 7-$\sigma$ W mass anomaly recently reported by CDF II collaboration. After applying all relevant constraints, including neutrino mass, W mass anomaly, and electroweak precision constraints, it is shown that the scalar triplet with hypercharge zero should have a mass in the range between $1.4~TeV$ and $5.19~TeV$, whereas the new fermion triplet can have a mass anywhere from $100$~TeV and up to $10^{13}$~GeV.
%
\begin{acknowledgments}
OP was supported by the Samsung Science and Technology Foundation under Grant No. SSTF-BA1602-04 and National Research Foundation of Korea under Grant Number 2018R1A2B6007000. RS is supported by
the Government of India, SERB Startup Grant SRG/2020/002303. All Feynman diagrams were created using TikZ-Feynman LateX package~\cite{Ellis:2016jkw}.
\end{acknowledgments}
\bibliography{references}
\end{document}